\documentclass[aps, prl,twocolumn]{revtex4}
\usepackage{color}
\usepackage{graphicx}
\usepackage{amsmath}
\usepackage{amssymb}
\usepackage{epstopdf}
\begin{document}
\title
{Microfluidization of graphite and formulation of graphene-based conductive inks}
\author{P. G. Karagiannidis$^{1}$, S. A. Hodge$^{1}$, L. Lombardi$^{1}$, F. Tomarchio$^{1}$, N. Decorde$^{1}$, S. Milana$^{1}$, I. Goykhman$^{1}$, Y. Su$^2$, S. V. Mesite$^2$, D. N. Johnstone$^3$, R. K. Leary$^3$, P. A. Midgley$^3$, N. M. Pugno$^{4,5,6}$, F. Torrisi$^{1}$, A. C. Ferrari$^{1}$}
\email{acf26@eng.cam.ac.uk}
\affiliation{$^1$Cambridge Graphene Centre, University of Cambridge, 9 JJ Thomson Avenue, Cambridge, CB3 0FA, UK\\
$^2$Microfluidics International Corporation 90 Glacier Dr., Suite 1000, Westwood, MA 02090, USA\\
$^3$Department of Materials Science \& Metallurgy, University of Cambridge, 27 Charles Babbage Road, Cambridge CB3 0FS, UK\\
$^4$Department of Civil, Environmental and Mechanical Engineering, University of Trento, Via Mesiano 77, 38123 Trento, Italy\\
$^5$Center for Materials and Microsystems, Fondazione Bruno Kessler, Via Sommarive 18, 38123 Povo, TN, Italy\\
$^6$School of Engineering and Materials Science, Queen Mary University of London, Mile End Road, London E1 4NS, UK}
\begin{abstract}
We report the exfoliation of graphite in aqueous solutions under high shear rate [$\sim10^8s^{-1}$] turbulent flow conditions, with a 100\% exfoliation yield. The material is stabilized without centrifugation at concentrations up to 100 g/L using carboxymethylcellulose sodium salt to formulate conductive printable inks. The sheet resistance of blade coated films is below$\sim2\Omega/\square$. This is a simple and scalable production route for graphene-based conductive inks for large area printing in flexible electronics.
\end{abstract}
\maketitle
\section{Introduction}
Printed electronics combines conducting, semiconducting and insulating materials with printing techniques, such as inkjet\cite{Caironi2013}, flexography\cite{Leppaniemi2015}, gravure\cite{Lau2013} and screen\cite{Krebs2010}. Metal inks based on Ag\cite{Dearden2005}, Cu\cite{Magdassi2010a} or Au\cite{Wang1997}, are used due to their high conductivity $\sigma\sim$10$^7$S/m\cite{Dearden2005,Jeong2008,Grouchko2011}. For flexible electronic devices, e.g. organic photovoltaics (OPVs), a sheet resistance, R$_S$ [=1/$\sigma$h, where h is the film thickness] $<$10$\Omega/\square$ is required\cite{Lucera2015}, while for printed radio-frequency identification (RFID) antennas one needs a few $\Omega/\square$\cite{Huang2015}. To minimize R$_S$ and cover the underneath rough layers such as printed poly(3,4-ethylenedioxythiophene) polystyrene sulfonate (PEDOT:PSS)\cite{Hosel2013}, thick films ($\mu$m range) are deposited using screen printing\cite{Hosel2013,SommerLarsen2013,Krebs2014,Caironi2015}. In this technique the ink is forced mechanically by a squeegee through the open areas of a stencil supported on a mesh of synthetic fabric\cite{Leach2007}. The ink must have high viscosity ($>$500mPas)\cite{Khan2015,Tobjork2011}, because lower viscosity inks run through the mesh rather than dispensing out of it\cite{Khan2015}. To achieve this viscosity, typical formulations of screen inks contain a conductive filler, such as Ag particles\cite{Merilampi2009}, and insulating additives\cite{Leach2007}, at a total concentration higher than C=100 g/L\cite{Leach2007}. Of this,$>$60g/L consist of the conductive filler needed to achieve high $\sigma\sim$10$^7$S/m\cite{Hyun2015a,Merilampi2009}. In 2016, the average cost of Ag was$\sim$550\$/Kg\cite{silverprice} and that of Au$\sim$40,000\$/Kg\cite{silverprice}, while the cost of graphite was$\sim$1\$/Kg\cite{statista}. However carbon/graphite inks are not typically used as printed electrodes in OPVs or RFIDs, due to their low $\sigma\sim$2-4x10$^3$S/m\cite{Gwent2015,Henkel2015,Dupont2015}, which corresponds to a R$_s\sim$20 to 10$\Omega/\square$ for a 25$\mu$m film. Cu inks can be used as a cheaper alternative (the 2016 cost of Cu was$\sim$4.7\$/Kg\cite{infomine}). However, metal electrodes can degrade the device performance, by chemically reacting with photoactive layers (Cu\cite{Lachkar1994}), by migrating into device layers (Cu\cite{Kim2011}, Ag\cite{Rosch2012}) or by oxidation (Ag\cite{Lloyd2009}). It is also reported that they might cause water toxicity\cite{Sondergaard2014}, cytotoxicity\cite{Fahmy2009}, genotoxicity\cite{Ahamed2010}, and deoxyribonucleic acid (DNA) damage\cite{Karlsson2008}. Thus, there is a need for cheap, stable and non-toxic conductive materials.

Graphene is a promising alternative conductive filler. Graphite can be exfoliated via sonication using solvents\cite{Hernandez2008,Valles2008,Khan2010, Hasan2010,Hernandez2010, Bourlinos2009} or water/surfactant solutions\cite{Lotya2009, Hasan2010}. Dispersions of single layer graphene (SLG) flakes can be produced at concentrations$\sim$0.01g/L\cite{Hernandez2008} with a yield by weight Y$_W\sim$1\%\cite{Hernandez2008}. Where, Y$_W$ is defined as the ratio between the weight of dispersed material and that of the starting graphite flakes\cite{Bonaccorso2012}. Dispersions of few layer graphene (FLG) ($<$4nm) can be achieved with C$\sim$0.1g/L\cite{Torrisi2012} in N-Methyl-2-pyrrolidone (NMP) and$\sim$0.2 g/L in water\cite{Hasan2010}. The low Y$_W\sim$1-2$\%$\cite{Hasan2010,Torrisi2012} for FLG in bath sonication is due to the fact that a significant amount of graphite remains unexfoliated as the ultrasonic intensity (i.e. the energy transmitted per unit time and unit area, J/cm$^2$s=W/cm$^2$\cite{Martinez2009}) is not uniformly applied in the bath\cite{Martinez2009,Nascentes2001} and depends on the design and location of the ultrasonic transducers\cite{Nascentes2001}. In tip sonication, the ultrasound intensity decays exponentially with distance from the tip\cite{Chivate1995}, and is dissipated at distances as low as$\sim$1cm\cite{Chivate1995}. Therefore, only a small volume near the tip is processed\cite{McClements2005}. Refs.\cite{Secor2013,Hyun2015b} reported$\sim$2nm thick flakes with lateral size $\sim$50-70x50-70nm$^2$ and C$\sim$0.2 g/L with Y$_W$=1\% by tip sonication. In order to formulate screen printing inks\cite{Hyun2015b}, the flakes C was increased from 0.2 g/L to 80 g/L via repetitive centrifugation (4 times) and re-dispersion (3 times) processes, resulting in an increased preparation time. Ref\cite{Paton2014} used a rotor-stator mixer to exfoliate graphite, reaching C$<$0.1g/L of FLGs with Y$_W<$2x$10^{-3}$\cite{Paton2014}. The low Y$_W$ is because in mixers, high shear rate, $\dot\gamma\sim$2x$10^4$-1x$10^5s^{-1}$ (i.e. the velocity gradient in a flowing material\cite{Brookfield}) is localized in the rotor stator gap\cite{Paul2004,Paton2014}, and can drop by a factor 100 outside it\cite{Paul2004}. Ref.\cite{Wang2011} reported the production of FLGs with number of layers, N$<$5 and Y$_W>$70\% through electrochemical expansion of graphite in lithium perchlorate/propylene carbonate. The process required 3 cycles of electrochemical charging followed by $>$10h of sonication and several washing steps (with hydrochloric acid/dimethylformamide, ammonia, water, isopropanol and tetrahydrofuran) to remove the salts. A method with less processing steps and high Y$_W$ (ideally 100\%) remains a challenge.

\begin{figure}
\centerline{\includegraphics[width=85mm]{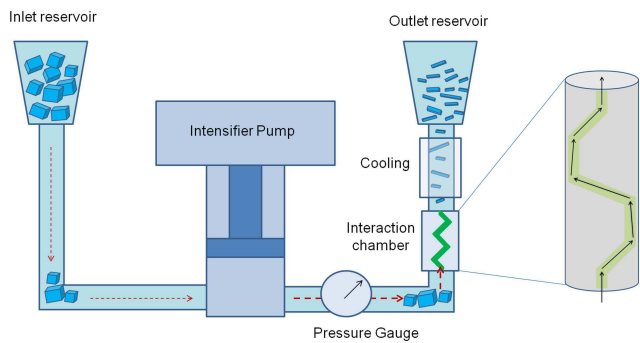}}
\caption{Schematic of the microfluidization process. Graphite flakes in SDC/water are added in the inlet reservoir. An intensifier pump applies high pressure (207MPa) and forces the suspension to pass through the microchannel of the interaction chamber where intense $\dot\gamma\sim$9.2x10$^7$s$^{-1}$ is generated. The processed material is cooled down and collected from the outlet reservoir. The process can be repeated several times.}
\label{micro}
\end{figure}

Microfluidization is a homogenization method that applies high pressure (up to 207MPa)\cite{Panagiotou2008b} to a fluid forcing it to pass through a microchannel (diameter, d$<$100$\mu$m), as shown in Fig.\ref{micro}, and discussed in Methods. The key advantage over sonication and shear-mixing is that high $\dot\gamma>10^6s^{-1}$\cite{Posch2008,microfluidicscorp} is applied to the whole fluid volume\cite{microfluidicscorp}, and not just locally. Microfluidization was used for the production of polymer nanosuspensions\cite{Panagiotou2008b}, in pharmaceutical applications to produce liposome nanoparticles with d$<$80nm to be used in eye drops for drug delivery to the posterior segment tissues of the eye\cite{Lajunen2014}, or to produce aspirin nanoemulsions\cite{Tang2013}, as well as in food applications for oil-in-water nanoemulsions\cite{Jafari2007}. Microfluidization was also used for the de-agglomeration and dispersion of carbon nanotubes\cite{Panagiotou2008}.

Here, we  report the production of FLGs by microfluidization. The dispersion is stabilized at a C up to 100 g/L using carboxymethylcellulose sodium salt (CMC) (C=10g/L), with Y$_W\sim$100\%. 4\% of the exfoliated material consists of FLGs ($<$4nm) and 96\% are flakes in the 4 to 70nm thickness range. The stabilized dispersion is used for blade coating and screen printing. R$_S$ of blade coated films after thermal annealing (300$^{\circ}$C-40 min) reaches 2$\Omega$/sq at 25 $\mu$m ($\sigma$=2x10$^4$S/m), suitable for electrodes in devices such as OPVs\cite{Lucera2015,Benatto2014}, organic thin-film transistors (OTFTs)\cite{Nisato2016} or RFIDs\cite{Huang2015}. The inks formulated here are deposited on glass and paper substrates using blade coating and screen printing to demonstrate the viability for these applications (OPVs, OTFTs, RFIDs).
\begin{figure*}
\centerline{\includegraphics[width=150mm]{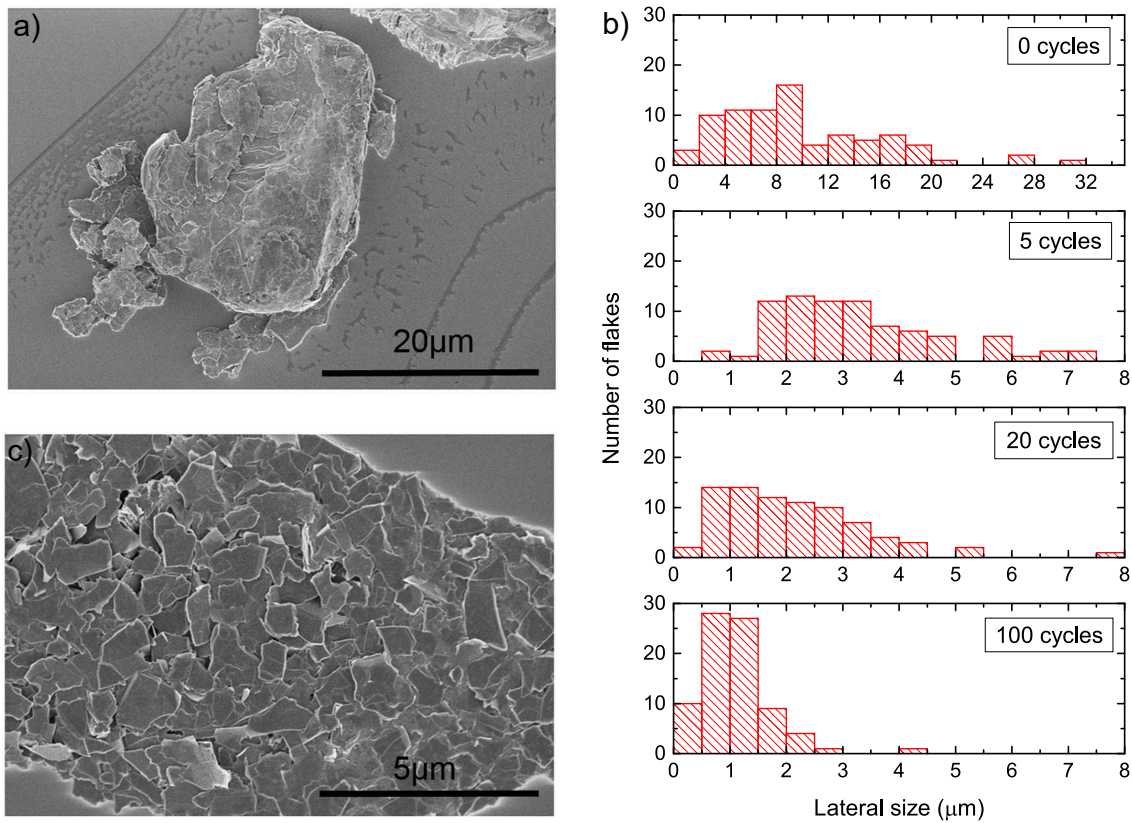}}
\caption{a) SEM image of pristine graphitic particles, b) histograms of lateral flake size for the starting material and after 5, 20 and 100 cycles, c) SEM image after 100 cycles.}
\label{sem1}
\end{figure*}
\begin{figure}
\centerline{\includegraphics[width=90mm]{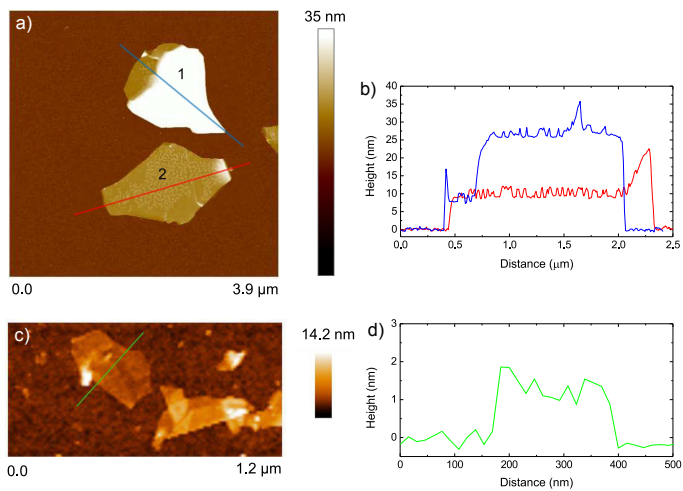}}
\caption{AFM images of typical flakes produced after 20 cycles: a) flakes with h=8.5 and 25nm. b) corresponding cross section profiles. c) flakes with h=1nm and d) corresponding cross section.}
\label{afm1}
\end{figure}
\begin{figure}
\centerline{\includegraphics[width=80mm]{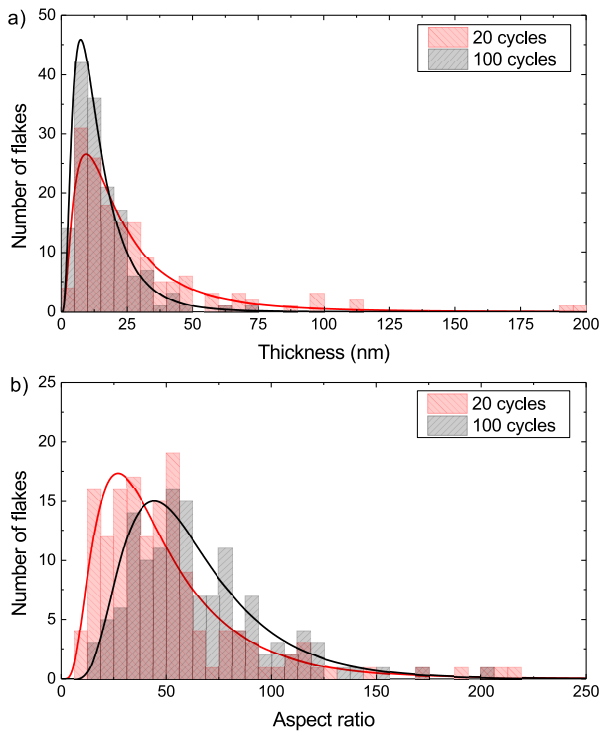}}
\caption{a) Flake thickness distribution and b) aspect ratio after 20 and 100 cycles, as measured by AFM.}
\label{afm2}
\end{figure}
\begin{figure*}
\centerline{\includegraphics[width=160mm]{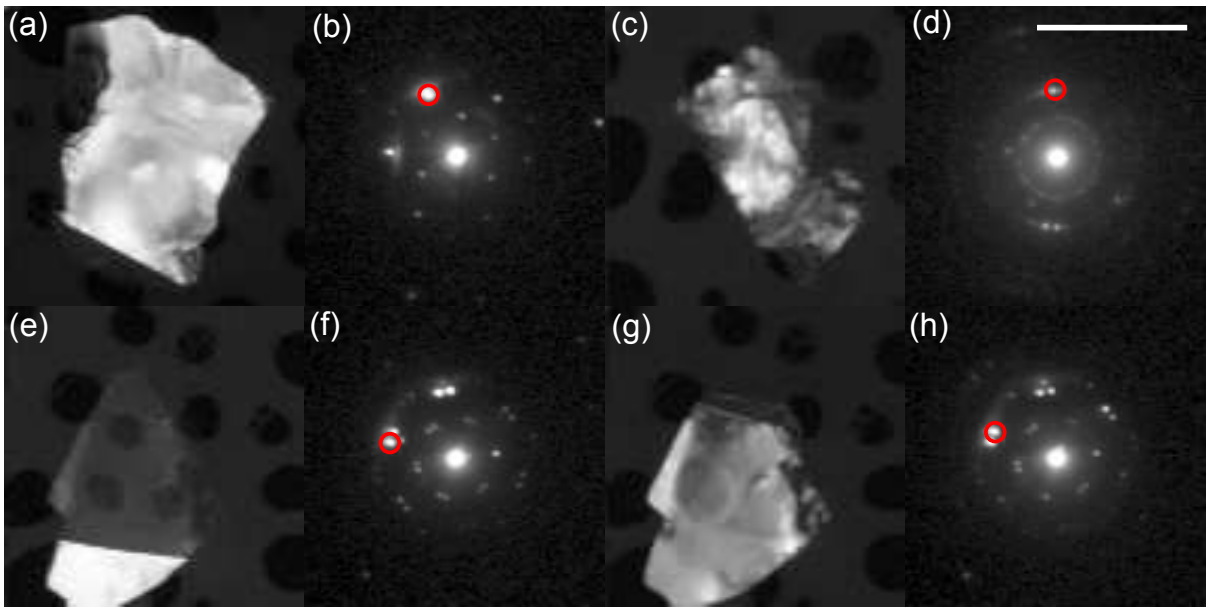}}
\caption{Virtual dark-field images (a,c,e,g) and representative diffraction patterns (b,d,f,h) acquired from (a,b) a single crystal flake, (c,d) a polycrystalline flake and (e-h) a polycrystalline flake comprising three crystals overlapping one another. The scale bar is 1$\mu$m.}
\label{TEM1}
\end{figure*}
\begin{figure*}
\centerline{\includegraphics[width=150mm]{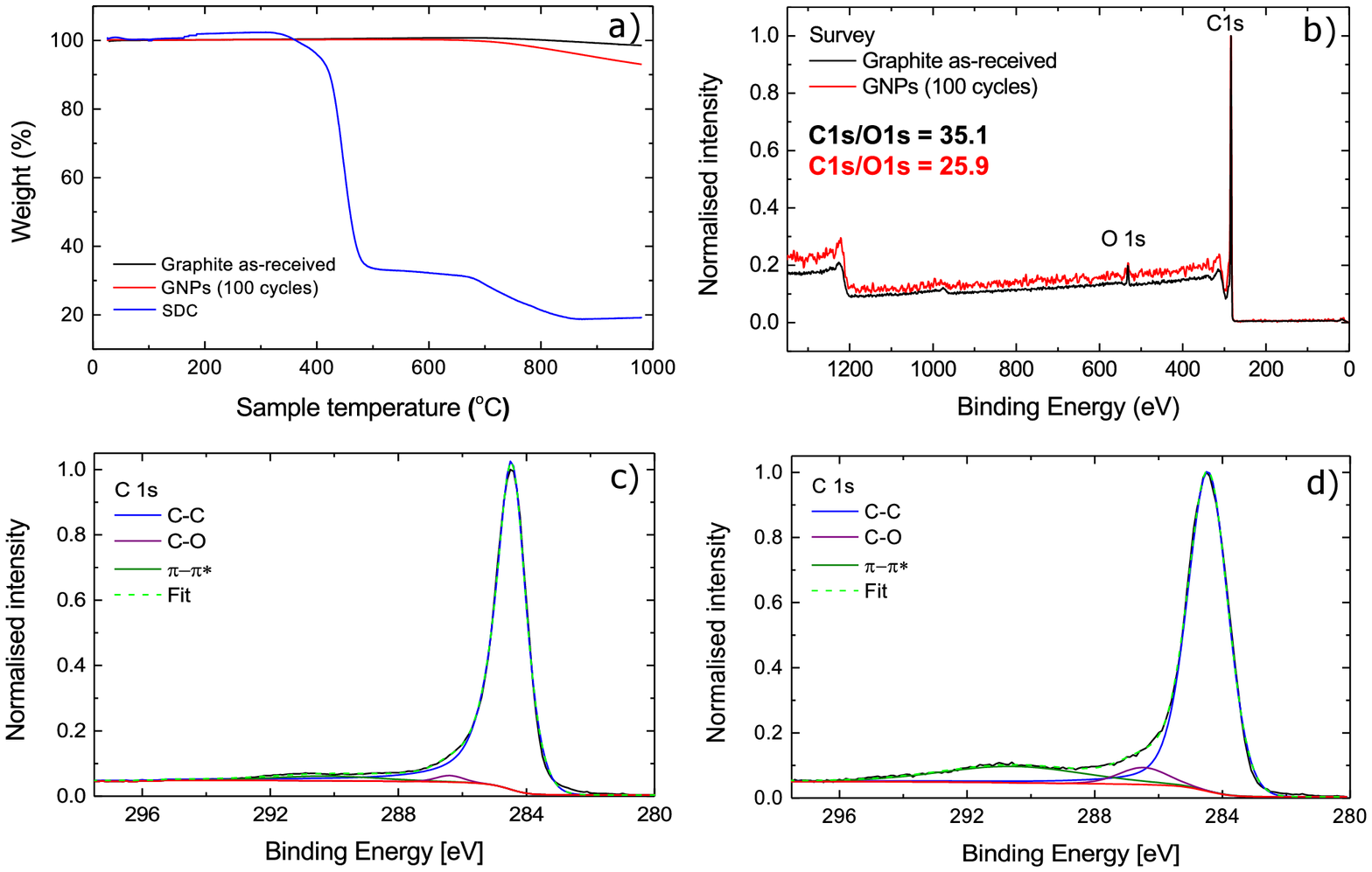}}
\caption{a) TGA of starting graphite, and flakes after 100 cycles and SDC in nitrogen. b) XPS of starting graphite and after 100 cycles. c)-d) high-resolution C1s spectra of starting graphite and after 100 cycles. Red curves represent the Shirley-type\cite{Shirley1972} background, which accounts for the effect of the inelastic scattering of electrons.}
\label{xps}
\end{figure*}
\section{Results and discussion}
We use Timrex KS25 graphite flakes as starting material. These are selected because their size (90\% are$<$27.2$\mu$m\cite{Imerys}) is suitable for flow in microchannels$\sim$87$\mu$m wide. Larger flakes would cause blockages. The flakes are used in conjunction with sodium deoxycholate (SDC) (Aldrich No.30970). They are first mixed in deionized (DI) water with SDC as a dispersing agent and then processed with a shear fluid processor (M-110P, Microfluidics International Corporation, Westwood, MA, USA) with a Z-type geometry interaction chamber, Fig\ref{micro}. Mixtures are processed at the maximum pressure which can be applied with this system (207MPa) with varying process cycles (1-100). The temperature, T [$^{\circ}$C]  increases from 20 to 55$^{\circ}$C after the liquid passes through the interaction chamber. A cooling system reduces it to$\sim$20$^{\circ}$C. This is important, otherwise T will keep increasing and the solvent will start to boil. Graphite/SDC mixtures with increasing graphite C (1-100g/L) and 9g/L SDC in DI water are processed over multiple cycles (1, 5, 10, 20, 30, 50, 70, 100). One cycle is defined as a complete pass of the mixture through the interaction chamber.

Scanning electron microscopy (SEM) (Fig.\ref{sem1}a) is used to assess the lateral size of the starting flakes and that of the exfoliated flakes after 5, 20 and 100 cycles. The exfoliated flakes are characterized as processed from the microfluidizer, with no centrifugation. Dispersions are diluted (1000 times, from 50g/L to 0.05 g/L) to avoid aggregation after they are drop cast onto Si/SiO$_2$. The samples are further washed with five drops of a mixture of water and ethanol (50:50 in volume) to remove the surfactant. Three different magnifications are used. For each magnification images are taken at 10 positions across each sample. A statistical analysis of over 80 particles (Fig.\ref{sem1}b) of the starting graphite reveals a lateral size (defined as the longest dimension) up to$\sim$32$\mu$m. Following microfluidization, this reduces, accompanied by a narrowing of the flake distribution. After 100 cycles (Fig.\ref{sem1}c), the mean flake size is$\sim$1$\mu$m.

Atomic force microscopy (AFM) is performed after 20 and 100 cycles to determine the thickness and aspect ratio (AR=lateral size/thickness). After 20 cycles, Fig.\ref{afm1}(a,b) shows flakes with d$\sim$1.7$\mu$m and h=25nm (flake 1) and d=1.9$\mu$m with h=8.5nm (flake 2). Fig.\ref{afm1}(c,d) shows$\sim$1nm thick flakes, consistent with N up to 3.

AFM statistics of flake thickness and AR are also performed. Three samples,$\sim$60$\mu$L, are collected from each dispersion (20 and 100 cycles) and drop cast onto 1cm x 1cm Si/SiO$_2$ substrates. These are further washed with five drops of a mixture of water and ethanol (50:50 in volume) to remove the surfactant. AFM scans are performed at 5 different locations on the substrate with each scan spanning an area of 20$\mu$mx20$\mu$m. For each processing condition we measure 150 flakes. After 20 cycles, h shows a lognormal distribution\cite{Kouroupis-Agalou2014} peaked at$\sim$10nm (Fig.\ref{afm2}a), with a mean value$\sim$19nm. After 100 cycles (Fig.\ref{afm2}a) the distribution is shifted towards lower h, with a maximum$\sim$7.4nm, a mean h$\sim$12nm (4\% of the flakes are $<$4nm and 96\% are between 4 and 70nm). Fig.\ref{afm2}b shows that AR increases with processing cycles. The mean AR increases from$\sim$41 for 20 cycles to$\sim$59 for 100 cycles.

The crystalline structure of individual flakes is investigated after 100 cycles (no statistical difference was observed between samples of different processing cycles) using scanning electron diffraction (SED)\cite{Moeck2011} performed on a Philips CM300 field emission gun transmission electron microscope (FEGTEM) operated at 50kV with a NanoMegas Digistar system\cite{nanomegas}. This enables the simultaneous scan and acquisition of electron diffraction patterns with an external optical CCD (charge-coupled device) camera imaging the phosphor viewing screen of the microscope. Using SED, small angle convergent beam electron diffraction patterns are acquired at every position as the electron beam is scanned over 10 flakes with a step size of 10.6nm.

Local crystallographic variations are visualized by plotting the diffracted intensity in a selected sub-set of pixels in each diffraction pattern as a function of probe position to form so-called "virtual dark-field" images\cite{Moeck2011,Gammer2015}. Fig.\ref{TEM1}a,c,e,g. shows the virtual dark-field images corresponding to the diffraction patterns in Fig.\ref{TEM1}b,d,f,h respectively. The virtual dark-field images show regions contributing to the selected Bragg reflection and therefore indicate local variations in the crystal structure and orientation. Consistent with selected area electron diffraction (SAED), three broad classes of flakes are observed, comprising (a,b) single crystals; (c,d) polycrystals with a small ($<$5) number of orientations, and (e-h) many ($>$5) small crystals. This shows that there is heterogeneity between individual flakes and that after 100 cycles a significant fraction ($\sim$70\%) are polycrystalline.

It is important to assess any chemical changes, such as oxidation or other covalent functionalization that might occur during processing since unwanted basal plane functionalisation may lead to a deterioration in electronic performance\cite{Punckt2013}. Flakes produced after 100 cycles are washed by filtration to remove SDC prior to thermogravimetric analysis (TGA) and X-ray photoelectron spectroscopy (XPS). For this washing procedure, 10mL ispopropanol is added to 5mL dispersion to precipitate the flakes. The resulting mixture is filtered through a 70mm diameter filter and rinsed with 500mL DI water followed by 500mL ethanol. The powder is dried under vacuum and scraped from the filter paper. Inert atmosphere (nitrogen) TGA is performed to identify adsorbed or covalently bonded functional groups using a TA Q50 (TA Instruments). Samples are heated from 25 to 100$^{\circ}$C at 10$^{\circ}$C/min, and then held isothermally at 100$^{\circ}$C for 10 min to remove residual moisture. T is then ramped to 1000$^{\circ}$C at a typical heating rate of 10$^{\circ}$C/min\cite{ASTM1}. The starting graphite shows$\sim$2wt\% decomposition above 700$^{\circ}$C. Flakes after washing reveal no surfactant, as confirmed by no weight loss at$\sim$400$^{\circ}$C where SDC suffers significant decomposition, as shown in Fig.6a. However, thermal decomposition of the flakes occurs at$\sim$600$^{\circ}$C, lower than the starting graphite, with a weight loss$\sim$6wt\%. Flakes with small lateral dimensions and thickness have a lower thermal stability compared to large area graphitic sheets\cite{Welham1998,Benson2014}.

The starting graphite and the exfoliated flakes are then fixed onto adhesive Cu tape for XPS\cite{ASTM2}. Unattached powder is removed by gently blowing with a nitrogen gun so as not to contaminate the ultra-high vacuum system. XPS is performed using an Escalab 250Xi (Thermo Scientific). The binding energies are adjusted to the sp$^{\text{2}}$ C1s peak of graphite at 284.5eV\cite{Moulder1992,Phaner-Goutorbe1994,Briggs1990}. Survey scan spectra (Fig.\ref{xps}b) of the starting graphite and the exfoliated flakes reveal only C1s and O1s$\sim$531eV\cite{Moulder1992} peaks. The slight increase in oxygen content for the exfoliated flakes compared to the starting material (C1s/O1s 35.1 to 25.9) is likely due to the increased ratio of edge to basal plane sites as the flake lateral size decreases following processing. However, C1s/O1s remains an order of magnitude larger than that typically observed in graphene oxide (GO) ($\sim$3\cite{Yang2009,Drewniak2015,Haubner2010}). Even following reductive treatments, the C1s/O1s ratio in reduced graphene oxide (RGO) does not exceed$\sim$15\cite{Yang2009, Drewniak2015}, i.e. half the ratio measured for our flakes. High-energy resolution (50eV pass energy) scans are then performed in order to deconvolute the C1s lineshapes. Both the starting graphite and exfoliated flakes can be fitted with 3 components (Fig.\ref{xps}c-d): an asymmetric sp$^{2}$ C-C (284.5eV\cite{Briggs1990,Moulder1992}), C-O ($\sim$285-286eV\cite{Briggs1990}) and $\pi$-$\pi$* transitions at$\sim$290eV\cite{Briggs1990}. Only a slight increase in the relative area of the C-O peak is seen (from $\sim$2\% in the starting graphite to $\sim$5\% in the exfoliated flakes). Therefore, we confirm that excessive oxidation or additional unwanted chemical functionalisations do not occur during microfluidization.
\begin{figure*}
\centerline{\includegraphics[width=180mm]{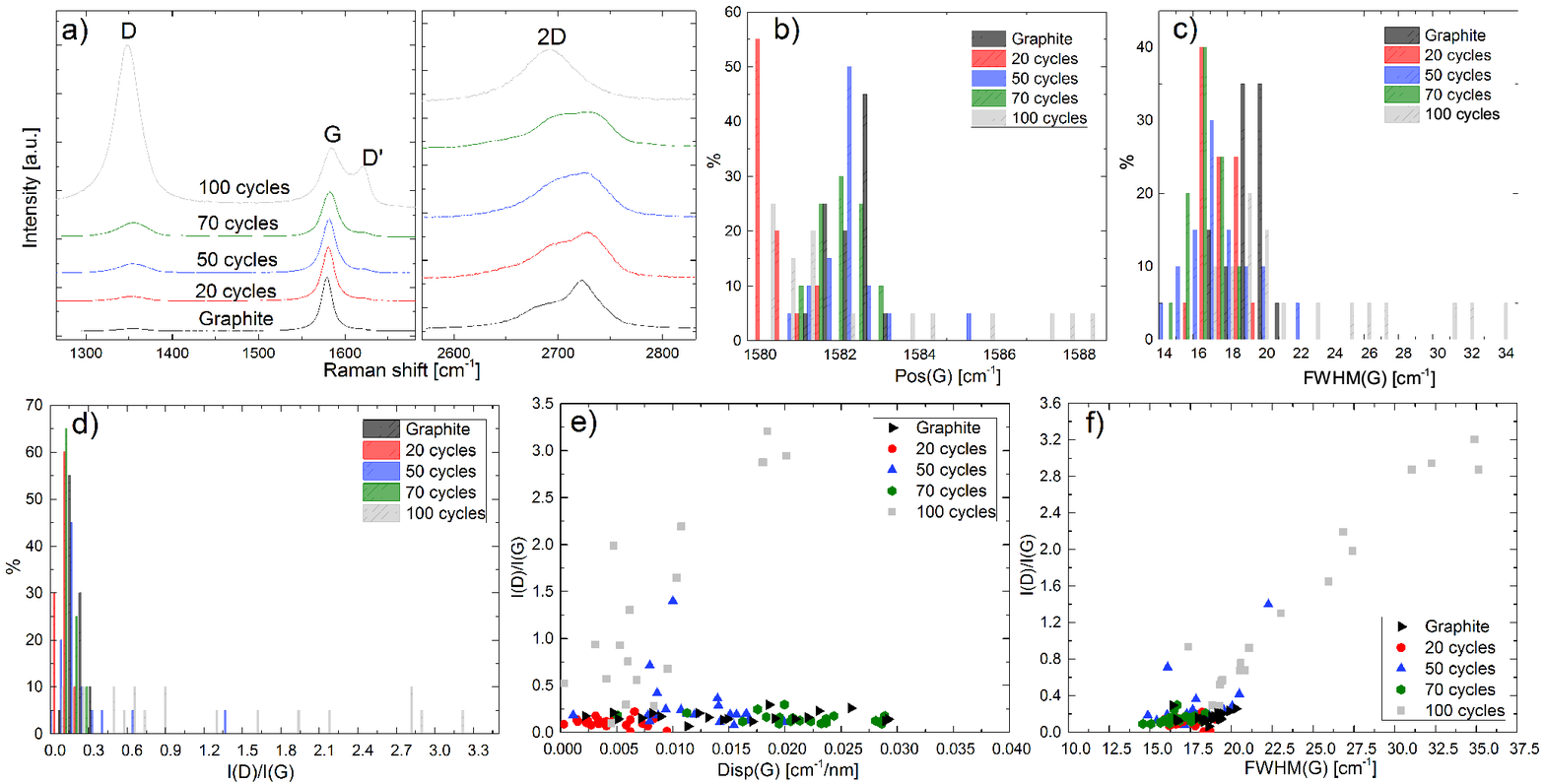}}
\caption{a) Raman spectrum at 514.5 nm for graphite and representative flakes after 20 (red curve), 50 (blue curve), 70 (green curve) and 100 (grey curve) process cycles, b) Distribution of Pos(G), c) FWHM(G), d) I(D)/I(G), and distribution of I(D)/I(G) as a function of e) Disp(G) and f) FWHM(G).}
\label{raman}
\end{figure*}

Raman spectroscopy is then used to assess the structural quality of the flakes. $\sim$60$\mu L$ of aqueous dispersion is drop cast onto 1cm x 1cm Si/SiO$_2$ substrates, then heated at 80-100 $^{\circ}$C for 20 min, to ensure water evaporation, and washed with a mixture of water and ethanol (50:50 in volume) to remove SDC. Raman spectra are acquired at 457, 514 and 633 nm using a Renishaw InVia spectrometer equipped with a 50x objective. The power on the sample is kept below 1mW to avoid any possible damage. The spectral resolution is$\sim$1cm$^{-1}$. A statistical analysis is performed on the samples processed for 20, 50, 70 and 100 cycles. The starting graphite powder is also measured. The Raman spectra are collected by using a motorized stage as follows: the substrate is divided in nine equally spaced regions of 200x200$\mu$m$^2$. In each region 3 points are acquired. This procedure is repeated for for each sample and for the 3 wavelengths. Statistical analysis is performed over 20 spectra in each of the 4 samples at the 3 different wavelengths.
\begin{figure*}
\centerline{\includegraphics[width=160mm]{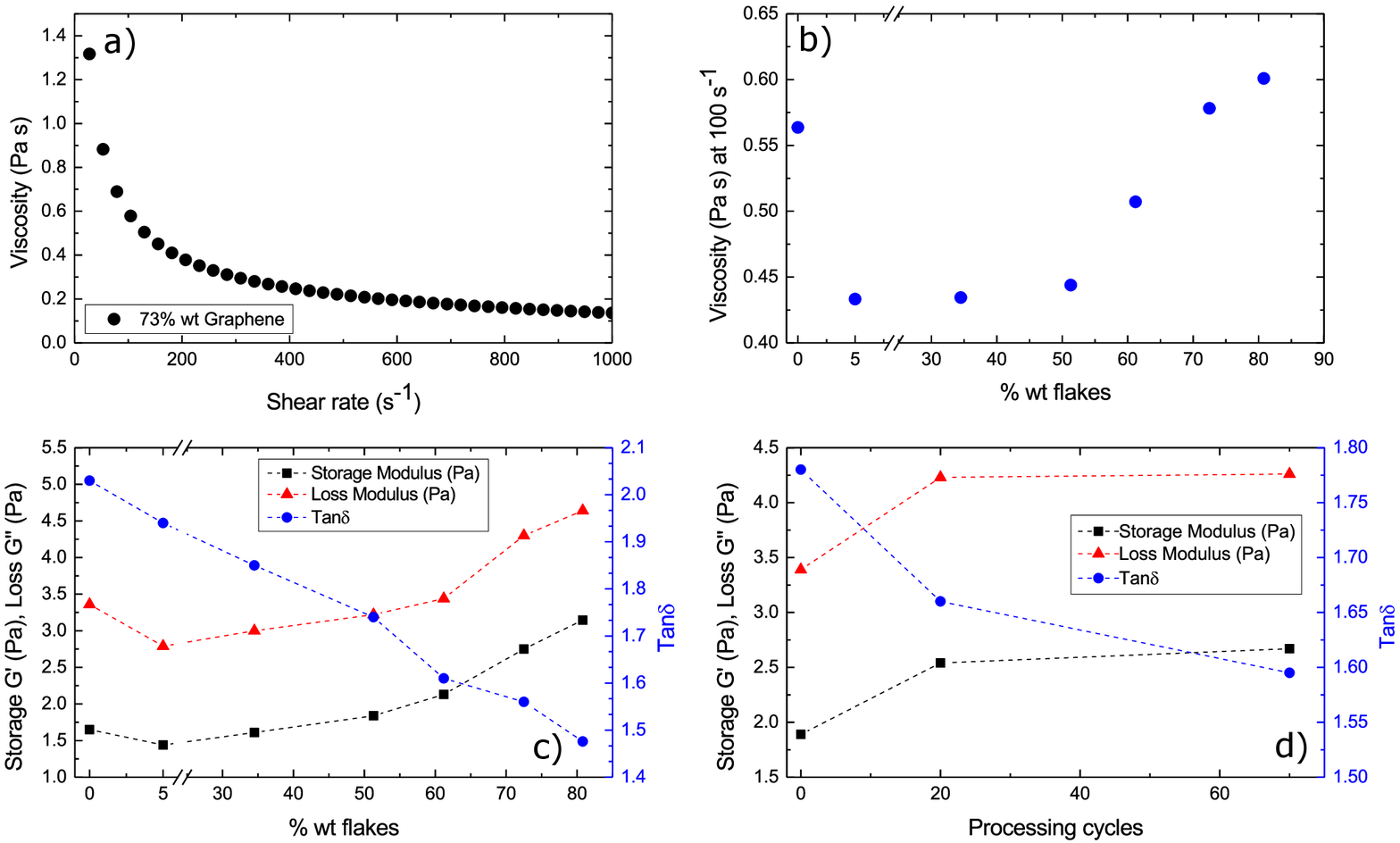}}
\caption{a) $\mu$ as a function of $\dot\gamma$ for an ink with 73wt\% flakes, b) $\mu$ at 100s$^{-1}$ for different wt\% of flakes. G',G'' and tan$\delta$ parameters as a function of c) wt\% flakes and d) processing cycles.}
\label{rheology}
\end{figure*}

The Raman spectrum of graphite has several characteristic peaks. The G peak corresponds to the high frequency E$_2g$  phonon at $\Gamma$\cite{Tuinstra1970}. The D peak is due to the breathing modes of six-atom rings and requires a defect for its activation\cite{Ferrari2000}. It comes from transverse optical (TO) phonons around the Brillouin zone corner K\cite{Tuinstra1970,Ferrari2000}. It is active by double resonance (DR)\cite{Thomsen2000,Baranov1987} and is strongly dispersive with excitation energy\cite{Pocsik1998} due to a Kohn Anomaly (KA) at K\cite{Piscanec2004}. DR can also happen as an intravalley process,i.e. connecting two points belonging to the same cone around K (or K'). This gives the so-called D' peak. The 2D peak is the D-peak overtone, and the 2D' peak is the D' overtone. Because the 2D and 2D' peaks originate from a process where momentum conservation is satisfied by two phonons with opposite wave vectors, no defects are required for their activation, and are thus always present\cite{Ferrari2006,Basko2009,Ferrari2013}. The 2D peak is a single Lorentzian in SLG, whereas it splits in several components as N increases, reflecting the evolution of the electronic band structure\cite{Ferrari2006}. In bulk graphite it consists of two components,$\sim$1/4 and 1/2 the height of the G peak\cite{Ferrari2006}. In disordered carbons, the position of the G peak, Pos(G), increases with decreasing of excitation wavelength (${\lambda_L}$)\cite{Ferrari2001}, resulting in a non-zero G peak dispersion, Disp(G) defined as the rate of change of Pos(G) with excitation wavelength. Disp(G) increases with disorder\cite{Ferrari2001}. Analogously to Disp(G), the full width at half maximum of the G peak, FWHM(G), increases with disorder\cite{Ferrari2003}. The analysis of the intensity ratio of the D to G peaks, I(D)/I(G), combined with that of FWHM(G) and Disp(G), allows one to discriminate between disorder localized at the edges and in the bulk. In the latter case, a higher I(D)/I(G) would correspond to higher FWHM(G) and Disp(G). Fig.\ref{raman}a plots representative spectra of the starting graphite (black line) and the processed flakes for 20 (red line), 50 (blue line), 70 (green line) and 100 cycles (grey line). The 2D band lineshape for the starting graphite and the 20-70 cycles samples shows two components (2D${_2}$,2D${_1}$). However, the intensity ratio I(2D${_2}$)/I (2D${_1}$) changes from$\sim$1.5 for starting graphite to$\sim$1.2 for 50 and 70 cycles until the 2D peak becomes a single component for 100 cycles, suggesting an evolution to electronically decoupled layers\cite{Ferrari2013,Ferrari2007}. FWHM(2D) for 100 cycles is$\sim$70cm$^{-1}$, significantly larger than in pristine graphene, but it is still a single Lorentzian. This implies that, even if the flakes are multilayers, they are electronically decoupled and, to a first approximation, behave as a collection of single layers. Pos(G) (Fig.\ref{raman}b), FWHM(G) (Fig.\ref{raman}c) and I(D)/I(G) (Fig.\ref{raman}d) for 20-70 cycles do not show a significant difference with respect to the starting graphite. However, for 100 cycles, Pos(G), FWHM(G) and I(D)/I(G) increase up to$\sim$1588, 34cm$^{-1}$ and 3.2 respectively suggesting a more disordered material. For all the processed samples (20-100) the D peak is present. For 20-70 cycles, it mostly arises from  edges, as supported by the absence of correlation between I(D)/I(G), Disp(G)(Fig.7e) and FWHM(G)(Fig.\ref{raman}f). Instead the correlation between I(D)/I(G), Disp(G)(Fig.7e) and FWHM(G)(Fig.\ref{raman}f) for 100 cycles indicates that D peak arises not only from edges, but also from in-plane defects. Therefore we select 70 cycles to formulate conductive printable inks.
\section{Printable inks formulation}
\begin{figure*}
\centerline{\includegraphics[width=160mm]{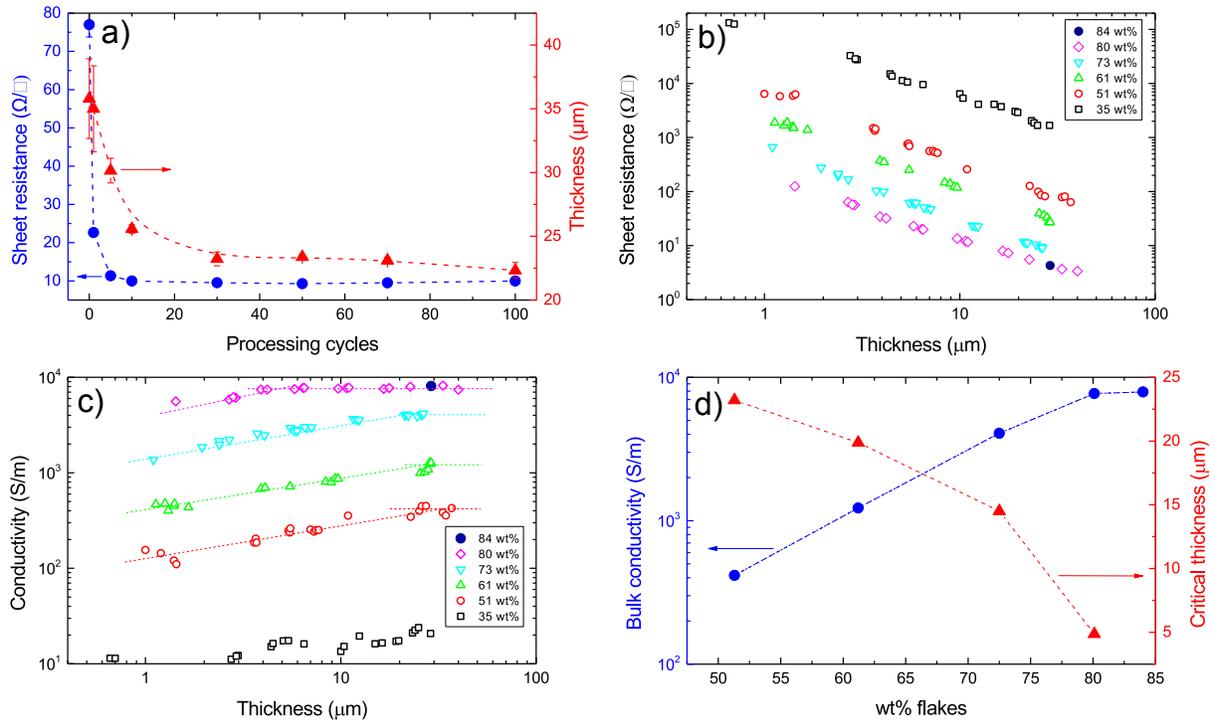}}
\caption{ a) R$_S$ and h as a function of processing cycles for a formulation with$\sim$73 wt\% flakes, b) R$_S$ as a function of h for different flake loadings (70 cycles), c) $\sigma$ as a function of h for different loadings, d) bulk $\sigma$ and critical h as a function of loading (70 cycles). All samples are dried for 10 min at 100$^{\circ}$C.}
\label{electrical}
\end{figure*}
Following microfluidization, carboxymethylcellulose sodium salt (CMC) (Weight Average Molecular Weight, M$_W$= 700.000, Aldrich No.419338), a biopolymer\cite{Ummartyotin2015} which is a rheology modifier\cite{Risio2007,Pavinatto2015}, is added to the dispersion to stabilize the flakes against sedimentation. CMC is added at C=10 g/L over a period of 3h at room temperature. This is necessary because if all CMC is added at once, aggregation occurs, and these aggregates are very difficult to dissolve. The mixture is continuously stirred until complete dissolution. Different inks are prepared keeping constant the SDC C=9 g/L and CMC C=10 g/L, while increasing the flakes C at 1, 10, 20, 30, 50, 80, 100 g/L. Once printed and dried, these formulations correspond to 5, 35, 51, 61, 73, 81 and 84 wt\% total solids content, respectively.

The rheological properties are investigated using a Discovery HR-1 rheometer from TA Instruments utilizing a parallel-plate (40mm diameter) setup\cite{Secco2014}. We monitor the elastic modulus G'[J/m$^3$=Pa]\cite{Mezger2006}, representing the elastic behavior of the material and a measure of the energy density stored by the material under a shear process\cite{Mezger2006}, and the loss modulus G''[J/m$^3$=Pa]\cite{Mezger2006}, representing the viscous behavior and a measure of the energy density lost during a shear process due to friction and internal motions\cite{Mezger2006}. Flow curves are measured by increasing $\dot\gamma$ from 1 to 1000$s^{-1}$ at a gap of 0.5mm, because this $\dot\gamma$ range is applied during screen printing\cite{Lin2008}. Fig.\ref{rheology}a plots the steady state viscosity of an ink containing 73\% wt flakes (100 process cycles) as a function of $\dot\gamma$. CMC imparts a drop in viscosity under shearing, from 570mPa.s at 100$s^{-1}$ to 140mPa.s at 1000$s^{-1}$. This is thixotropic behavior\cite{Benchabane2008}, since the viscosity reduces with $\dot\gamma$. The higher $\dot\gamma$, the lower the viscosity\cite{Benchabane2008}. This behavior is shown by some non-Newtonian fluids, such as polymer solutions\cite{Nijenhuis2007} and biological fluids\cite{Irgens2014}. It is caused by the disentanglement of polymer coils or increased orientation of polymer coils in the direction of the flow\cite{Benchabane2008}. On the other hand, in Newtonian liquids the viscosity does not change with $\dot\gamma$\cite{Irgens2014}. Refs.\cite{deButts1957,Elliot1974} reported that thixotropy in CMC solutions arises from the presence of unsubstituted (free) OH groups. Thixotropy decreases as the number of OH groups increases\cite{deButts1957,Elliot1974}.
\begin{figure}[b!]
\centerline{\includegraphics[width=85mm]{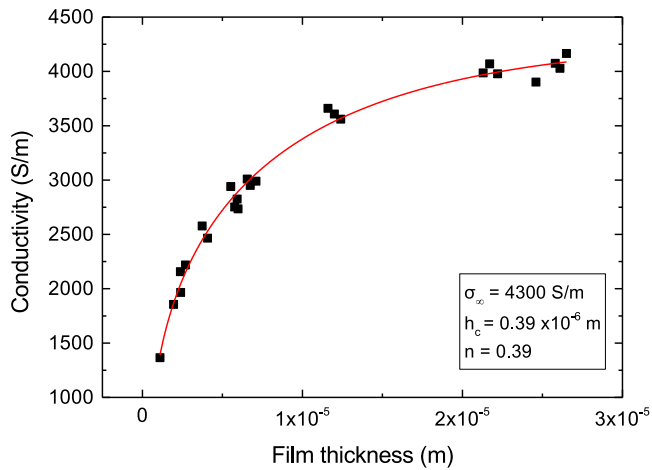}}
\caption{Fit of $\sigma$ as a function of h according to Eq.\ref{eq:sigma2} for 73 wt\% flakes.}
\label{fit}
\end{figure}
\begin{figure*}
\centerline{\includegraphics[width=180mm]{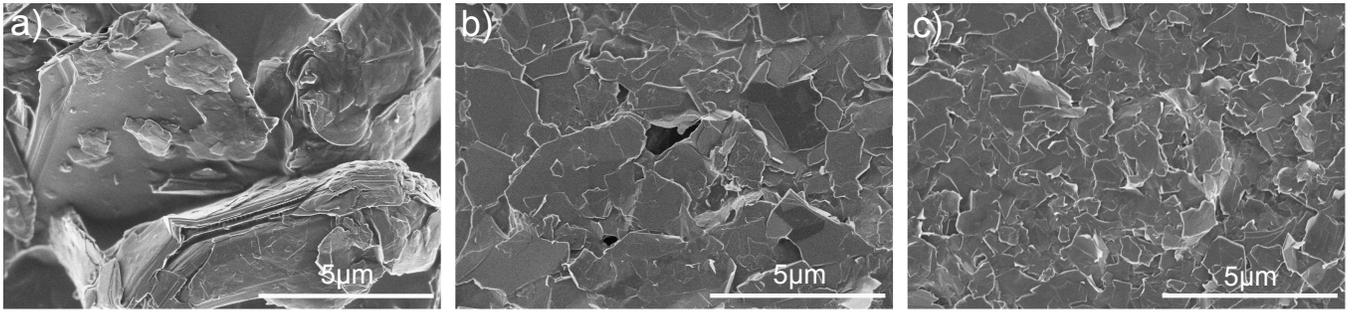}}
\caption{SEM images taken from coatings comprising a) starting graphite, b) after 5 cycles and c) after 100 cycles. The scale bar is 5$\mu$m.}
\label{sem2}
\end{figure*}

During printing, shear is applied to the ink and its viscosity decreases, making the ink easier to print or coat. This shear thinning behavior facilitates the use of the ink in techniques such as screen printing, in which a maximum $\dot\gamma\sim$1000s$^{-1}$ is reached when the ink is penetrating the screen mesh\cite{Lin2008}. Fig.\ref{rheology}b  plots the viscosity at 100s$^{-1}$ as a function of wt\% flakes (70 process cycles). This drops from 0.56 to 0.43Pa.s with the addition of 5 wt\% flakes, and recovers above 50 wt\% flakes. The CMC polymer (10 g/L in water) has a $\mu\sim$0.56Pa s at 100s$^{-1}$, and drops to 0.43Pa.s with the addition of 5 wt\% flakes. The loading wt\% of flakes affects $\dot\gamma$, which increases at 51 wt\% and reaches 0.6Pa s at 80 wt\%.

More information on the ink rheological behavior and microstructure can by obtained by oscillatory rheology measurements\cite{Clasen2001}. CMC gives a viscoelastic character to the ink. This can also be evaluated in terms of the loss factor defined as tan$\delta$=G''/G'\cite{Mezger2006}. The lower tan$\delta$, the more solid-like (i.e. elastic) the material at a given strain or frequency\cite{Mezger2006}. Fig.\ref{rheology}c plots G', G'' and tan$\delta$ at 1\% strain and frequency, checked from dynamic amplitude sweeps in order to be within the linear viscoelastic region (LVR). In LVR, G' and G'' are not stress or strain dependent\cite{Steffe1996} as a function of flake loading. Addition of 5 wt\% flakes in CMC decreases both G' and G'', which start to increase for loadings above 30 wt\%. Tan$\delta$ decreases with flake loading, leading to a more solid-like behavior. We estimate G', G'' and tan$\delta$ also for inks containing flakes processed at different cycles, while keeping the flakes loading at$\sim$73\%, Fig.\ref{rheology}d. Both G' and G'' increase with processing cycles, while tan$\delta$ decreases, indicating an increase of elastic behavior with processing.

Inks are blade coated onto glass microscope slides (25x75mm) using a spacer to control h. The films are dried at 100$^{\circ}$C for 10 min to remove water. h depends on the wet film thickness, the total solid content wt\% of the ink and the number of processing cycles. We investigate the effects of processing cycles, flake content and post-deposition annealing on R$_S$. This is measured in 4 different locations per sample using a four-point probe. A profilometer (DektakXT, Bruker) is used to determine h for each point. In order to test the effect of the processing cycles, films are prepared from inks containing$\sim$73wt\% flakes processed for 0, 5, 10, 30, 50, 70 and 100 cycles keeping the wet h constant (1mm). Fig.\ref{electrical}a shows the effect of processing cycles on R$_S$ and h. Without any processing, the films have R$_S\sim$77$\Omega/\square$ and h=35.8$\mu$m, corresponding to $\sigma\sim$3.6x$10^2$S/m. Microfluidization causes a drop in R$_S$ and h. 10 cycles are enough to reach$\sim$10$\Omega/\square$ and h$\sim$25.6$\mu$m, corresponding to $\sigma\sim$3.9x$10^3$S/m. R$_S$ does not change significantly between 10 and 100 cycles, while h slightly decreases. We get$\sigma\sim$4.5x$10^2$S/m above 30 cycles.
\begin{figure*}
\centerline{\includegraphics[width=160mm]{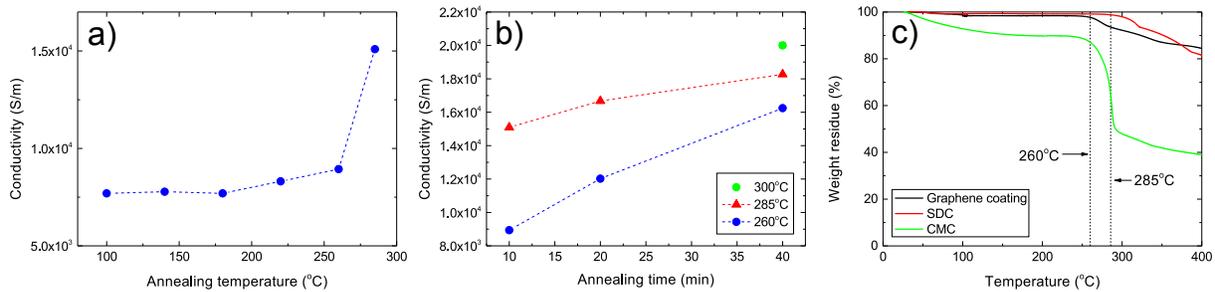}}
\caption{$\sigma$ as a function of a) annealing T and b) time. c) TGA thermograms from composite coatings along with the SDC (powder) and the CMC (powder) components.}
\label{electrical2}
\end{figure*}
\begin{figure*}
\centerline{\includegraphics[width=160mm]{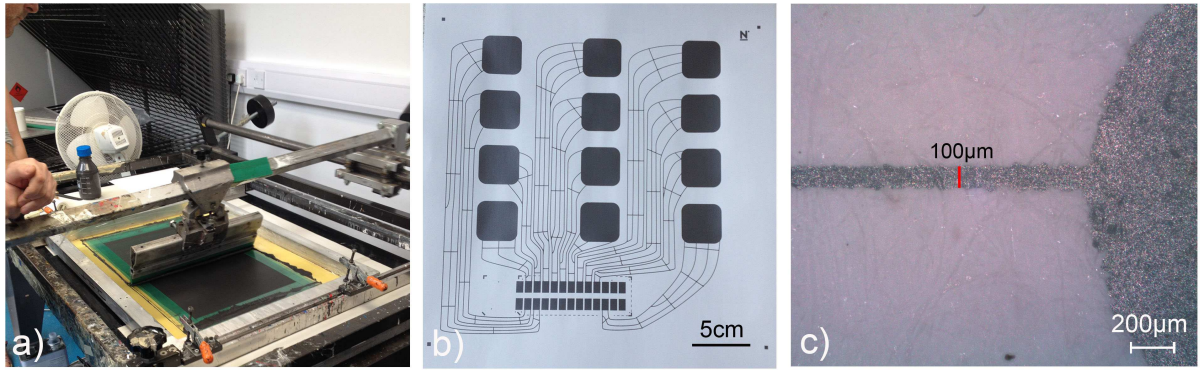}}
\caption{a) Demonstration of screen printing, b) capacitive touchpad design (29cmx29cm) printed on paper, c) the line resolution is 100$\mu$m.}
\label{Printing}
\end{figure*}

The effect of flake loading at fixed processing cycles (70 cycles) is investigated as follows. Dispersions with different loadings are prepared by increasing the flakes C between 1 and 100g/L, whilst keeping the SDC (9g/L) and CMC (10g/L) constant. Films of different h are prepared by changing the spacer height during blade coating, leading to different wet and dry h. R$_S$ and $\sigma$ as a function of h are shown in Figs.\ref{electrical}b,c. At$\sim$34.5wt\% the flakes already form a percolative network within the CMC matrix and $\sigma\sim$15-20S/m is achieved ($\sigma$ of cellulose derivative films is$<10^{-8}$S/m\cite{Roff1971}). Fig.\ref{electrical}c shows that, for a given composition, there is a critical h below which $\sigma$ is thickness dependent. Above this h, the bulk $\sigma$ is reached. As shown in Fig.\ref{electrical}c, for a loading$\sim$80wt\% we get $\sigma\sim$7.7x$10^{3}$S/m for h$>$4.5$\mu$m. Higher loadings (84 wt\%) do not increase $\sigma$ further. Fig.\ref{electrical}d indicates that the critical h where the bulk $\sigma$ is reached drops from$\sim$20$\mu$m for 51 wt\% to $\sim$4.5$\mu$m for 80 wt\%. Coatings with h$>$4.5$\mu$m can be easily achieved using screen printing in a single printing pass.

Fig.\ref{electrical}c shows that $\sigma$ is h dependent up to a critical point. In order to understand an effect of h on $\sigma$ we adapt the percolation model based on Ref.\cite{Xia1988}. The total area covered by non-overlapping flakes is A$_f$ (e.g. for elliptical flakes A$_f$=m$\pi$ab where m is the number of flakes and a [m] and b [m] are their half-axes lengths). The fractional area covered by the flakes (overlapping), with respect to the total area S[m$^2$], can be evaluated as q=1-p, with p=e$^{-A_f/S}$ where q is the fractional area covered by the flakes\cite{Xia1988}. q coincides with A$_f$/S only when the flakes do not overlap. Denoting by A$_f$h$_f$ the total flakes volume and f the volume fraction of flakes in the films we have:
\begin{equation} \label{eq:Af}
A_f h_f=fhS=-Sh_flnp
\end{equation}
$\sigma$ follows a power law behavior of the form of\cite{Xia1988}:
\begin{equation} \label{eq:sigma1}
 \sigma=k(q-q_c)^n
\end{equation}
around the percolation threshold q$_{\text{c}}$\cite{Xia1988}, and n is the electrical conductivity critical exponent above percolation. Eqs.\ref{eq:Af},\ref{eq:sigma1} give the following:
\begin{equation} \label{eq:sigma2}
\sigma = \sigma_\infty \Big[1-e^\frac{(h_c-h) f}{\ h_f}\Big]^n
\end{equation}
where $\sigma_\infty$=\emph{k}e($^{-fnh_c/h_f}$) and h$_c$ is the critical thickness corresponding to zero $\sigma$. $\sigma$ as a function of h is fitted with Eq.\ref{eq:sigma2} in Fig.\ref{fit} for a formulation containing$\sim$73wt\% flakes, i.e. f=0.61. From the fit we get $\sigma_\infty\sim$4.3x10$^3$S/m, h$_c$=0.39$\mu$m, h$_f\sim$7.58$\mu$m and n=0.39.

Fig.\ref{sem2}, shows SEM images of the coatings comprising the starting graphite (Fig.\ref{sem2}a), after 5 (Fig.\ref{sem2}b) and 100 cycles (Fig.\ref{sem2}c). Flake size reduction and platelet-like morphology is observed after microfluidic processing. The samples have fewer voids compared to the starting graphite, providing higher interparticle contact area and higher flakes packing density consistent with the reduction of h (Fig.\ref{electrical}a) and an increased $\sigma$. Whilst the density increase results in more pathways for conduction, the smaller flake size increases the number of inter-particle contacts. Subsequently, R$_S$ remains constant.

The effect of post-deposition annealing is studied using blade coated films for a formulation containing$\sim$80wt\% flakes after 70 cycles. After drying, the films are annealed for 10 min between 100 and 285$^{\circ}$C. Fig.\ref{electrical2}a plots $\sigma$ as a function of T from 100 to 285$^{\circ}$C. A three step regime can be seen. In the first (100-180$^{\circ}$C) $\sigma$ is constant ($\sim$7.7x10$^3$S/m), above 180$^{\circ}$C it increases, reaching 9x10$^3$S/m at 260$^{\circ}$C and a significant increase is observed at 285$^{\circ}$C to$\sim$1.5x$10^4$S/m. Fig.\ref{electrical2}b shows the effect of annealing time at 260, 285 or 300$^{\circ}$C. Either higher T or longer annealing times are required to increase $\sigma$.

TGA is then used to investigate the thermal stability of the films, Fig.\ref{electrical2}c. The thermogram of CMC polymer reveals a 10\% weight loss up to 200$^{\circ}$C, due to water loss\cite{Li2009}. Fig.\ref{electrical2} also shows that 50\% of the CMC is decomposed at 285$^{\circ}$C, while the SDC surfactant remains intact. Annealing at 300$^{\circ}$C for 40min leads to films with R$_S\sim2\Omega/\square$ (25$\mu$m) corresponding to $\sigma\sim$2x10$^4$S/m. This $\sigma$ is remarkable, given the absence of the centrifugation step that is usually performed to remove the non-exfoliated material or washing steps to remove the non-conductive polymer and surfactant materials.

The printability of the ink with$\sim$80wt\% flakes after 70 cycles is tested using a semi-automatic flatbed screen printer (Kippax kpx 2012) and a Natgraph screen printer (Fig\ref{Printing}a), both equipped with screens with 120 mesh count per inch. Trials are made onto paper substrates. A Nikon optical microscope (Eclipse LV100) is used to check the printed patterns. Fig.\ref{Printing}b shows a 29 cm x 29 cm print on paper with a line resolution$\sim$100$\mu$m (Fig.\ref{Printing}c). The printed pattern (Fig.\ref{Printing}b) can be used as a capacitive touch pad in a sound platform that translates touch into audio\cite{audioposter}.
\section{Conclusions}
We reported a simple and scalable route to exfoliate graphite. The resulting material can be used without any additional steps (washing or centrifugation) to formulate highly conductive inks with adjustable viscosity for high throughput printing. Conductivity as high as 2x10$^4$ S/m was demonstrated. Our approach enables the mass production of chemically unmodified flakes that can be used in inks, coatings and conductive composites for a wide range of applications.
\section{Acknowledgements}
We acknowledge funding from EU Graphene Flagship, ERCs Grants Hetero2D, HiGRAPHINK, 3DIMAGEEPSRC, ESTEEM2, BIHSNAM, KNOTOUGH, SILKENE, EPSRC Grants EP/K01711X/1, EP/K017144/1,a Vice Chancellor award from the University of Cambridge, a Junior Research Fellowship from Clare College and the Cambridge NanoCDT and GrapheneCDT. We acknowledge Chris Jones for useful discussions, and Imerys Graphite \& Carbon Switzerland Ltd. for providing graphite powders.
\section{Methods}
\section{Microfluidization process}
In order to compare the microfluidization process with sonication or shear mixing it is important to elucidate the chamber fluid dynamics in the microfluidizer. The mean channel velocity U [m/s] of the fluid inside the microchannel is\cite{Munson2009}:
\begin{equation} \label{eq:U}
U=\frac{Q}{A}
\end{equation}
where Q [m$^3$/s] is the volumetric flow rate, defined as\cite{Rouse1946}:
\begin{equation} \label{eq:Q}
Q=\frac{c_n V}{t}
\end{equation}
where c$_n$ is the number of cycles, V[m$^3$] the volume of material (graphite and solvent) passing a point in the pipe per unit time t[s] and A[m$^2$] is the cross-sectional area of the pipe given by:
\begin{equation} \label{eq:A}
\ A = \pi \left(\frac{D_h}{2}\right)^2
\end{equation}
where D$_h$[m] is the hydraulic diameter of the microchannel, defined as four times the ratio of the cross-sectional flow area divided by the wetted perimeter (the part of the microchannel in contact with the flowing fluid), P, (D$_h$=4A/P)\cite{Munson2009}. For a batch of 0.18L it takes 1.93h to complete 70 cycles. Thus Eq.\ref{eq:U} gives Q=1.8x$10^{-6}$m$^3$/s. Eq.\ref{eq:A} with D$_h\sim$87$\mu$m\cite{microfluidicscorp} gives A=5940x$10^{-12}$m$^2$. Thus, Eq.\ref{eq:U} gives U$\sim$304m/s.

The Reynolds number, Re, can be used to determine the type of flow inside a microchannel\cite{Munson2009} and it is given by\cite{Munson2009}:
\begin{equation} \label{eq:Re}
Re=\frac{\rho U D_h}{\mu}
\end{equation}
where $\rho$[Kg/m$^3$] is the liquid density. We typically use 50 up to 100g/L of graphite, which corresponds to a total density (mixture of graphite and water) of 1026 to 1052Kg$/$m$^3$. $\mu$[Pa s] is the dynamic viscosity ($\mu$=$\tau$/$\dot\gamma$, where $\tau$[Pa] is the shear stress). We measure $\mu$ with a rotational rheometer in which a known $\dot\gamma$ is applied to the sample and the resultant  torque (or $\tau$) is measured\cite{Secco2014}. We get $\mu\sim$1x$10^{-3}$ Pa s (20$^{\circ}$C), similar to water\cite{Munson2009}. Thus, Eq.\ref{eq:Re} gives Re$\sim$2.7x10$^4$, which indicates that there is a fully developed turbulent flow inside the microchannel (there is a transition from laminar to turbulent flow in the 2000$>$Re$>$4000 range)\cite{Holman1986}.

The pressure losses inside the channel can be estimated by the Darcy-Weisbach equation\cite{Munson2009}, which relates the pressure drop, due to friction along a given length of pipe, to the average velocity of the fluid flow for an incompressible fluid\cite{Munson2009}:
\begin{equation} \label{eq:Dp}
\Delta p = \frac{ f_D L \rho U^2}{\ 2D}
\end{equation}
where $\Delta p$ [Pa] is the pressure drop, L[m] is the pipe length, f$_D$ is the Darcy friction factor which is a dimensionless quantity used for the description of friction losses in pipe flow\cite{Munson2009}. For Re=2.7x10$^4$, f$_D$ $\sim$0.052, as obtained from the Moody chart\cite{Moody1944}, which links f$_D$, Re, and relative roughness of the pipe (=absolute roughness/hydraulic diameter\cite{Munson2009}). From Eq.\ref{eq:Dp} we get $\Delta p\sim$3.25x$10^7$Pa. The energy dissipation rate per unit mass $\epsilon$ [m$^2$/s$^3$] inside the channel can be written as\cite{Siddiqui2009}:
\begin{equation} \label{eq:e}
\epsilon = \frac{\ Q \Delta p}{\rho V}
\end{equation}
From Eqs.\ref{eq:Q} and \ref{eq:Dp} we get $\epsilon\sim$8.3x$10^9$ m$^2$/s$^3$. $\dot\gamma$ can then be estimated as\cite{Boxall2012}:
\begin{equation} \label{eq:gamma}
\dot{\gamma}=\sqrt \frac{\epsilon}{\nu}
\end{equation}
where $\nu$[m$^2$/s] is the kinematic viscosity\cite{Boxall2012}, defined as $\nu=\mu/\rho\sim$1x$10^{-6}m^2$/s. From Eq.\ref{eq:gamma} we get $\dot{\gamma}\sim10^8s^{-1}$, which is 4 orders of magnitude higher than the $\dot\gamma$ required to initiate graphite exfoliation\cite{Paton2014}. Thus, the exfoliation in the microfluidizer is primarily due to shear and stress generated by the turbulent flow. In comparison, in a rotor-stator shear mixer, lower $\dot{\gamma}\sim$2x$10^4$-1x$10^5s^{-1}$\cite{Zhang2012,Paul2004,Boxall2012}) are achieved and only near the probe of the rotor stator\cite{Paul2004}. Thus, exfoliation does not take place in the entire batch uniformly\cite{Paton2014}. On the contrary, in a microfluidizer all the material is uniformly exposed to high shear forces\cite{Panagiotou2008}.

Turbulent mixing is characterized by a near dissipationless cascade of energy\cite{Boxall2012}, i.e. the energy is transferred from large (on the order of the size of the flow geometry considered) random, three-dimensional eddy type motions to smaller ones (on the order of the size of a fluid particle)\cite{Munson2009}. This takes place from the inertial subrange (IS) of turbulence where inertial stresses dominate over viscous stresses, down to the Kolmogorov length\cite{Kolmogorov1941}, $\eta$[m], i.e. the length-scale above which the system is in the inertial subrange of turbulence, and below which it is in the viscous subrange (VS), where turbulence energy is dissipated by heat\cite{Richardson1922, Boxall2012}. $\eta$ can be calculated as\cite{Kolmogorov1941}:
\begin{equation} \label{eq:eta}
\eta=\left(\frac{\nu^3}{\epsilon}\right)^{\frac{1}{4}}
\end{equation}
From $\nu\sim$1x$10^{-6}m^2$/s (calculated above) and Eq.\ref{eq:e} , we get $\eta\sim$103nm for microfluidization in water. Since our starting graphitic particles are much larger ($>\mu$m) than $\eta$, exfoliation occurs in the IS of turbulence rather than VS, where energy is dissipated through viscous losses. In comparison, in a kitchen blender $\eta$=6$\mu$m\cite{Varrla2014}, thus exfoliation occurs in the VS, i.e. the energy is dissipated through viscous losses, rather than through particle disruption. During microfluidization, in the IS, the main stress contributing to exfoliation is due to pressure fluctuations, i.e. the graphite is bombarded with turbulent eddies. This stress,$\tau_{IS}$[Pa], can be estimated as\cite{Boxall2012}:
\begin{equation} \label{eq:tauis}
\tau_{\text{IS}}\sim\rho({\epsilon d_{\text{g}})^{\frac{2}{3}}}
\end{equation}
where d$_g$ is the diameter of a sphere of equivalent volume to the flakes. For d$_g$=0.1 to 27$\mu$m, $\tau_{IS}$ is in the range$\sim$0.1-4MPa. The dynamic pressure also breaks the flakes, as well as exfoliating them. For length scales$<\eta$, the flakes are in the VS and the stress applied on the flakes $\tau_{VS}$ can be estimated as\cite{Boxall2012}:
\begin{equation} \label{eq:tauvs}
\tau_{\text{VS}}\sim\mu \sqrt{{\frac{\epsilon}{\nu}}}
\end{equation}
which gives $\tau_{VS}\sim$0.1MPa. Thus, the stresses applied on the flakes in the IS are much higher than in the VS, where energy is lost by heat.

In microfluidization, the energy density E/V[J/m$^3$], (where E[J] is the energy) equates the pressure differential\cite{Jafari2007}, due to very short residence times $\sim$10$s^{-4}$\cite{Jafari2007}, i.e. the time the liquid spends in the microchannel. Therefore, for a processing pressure$\sim$207MPa, E/V=207MPa=2.07x10$^8$J/$m^3$. For this total energy input per unit volume, the flakes production rate P$_r$=VC/t [g/h] for a typical batch of V=0.18L and t=1.93h (for 70 cycles), is P$_r\sim$9.3g/h, with starting graphite concentration$\sim$100 g/L using a lab-scale system. Scaling up of the microfluidization process can be achieved by increasing Q, using a number of parallel microchannels\cite{microfluidicscorp}, which decreases the time required to process a given V and c$_n$ (Eq.\ref{eq:Q}). With shorter time, P$_r$ increases. Large scale microfluidizers can achieve flow rates$\sim$12L/min\cite{microfluidicscorp} at processing pressure$\sim$207MPa, which correspond to P$_r$=CQ$/$c$_n\sim$1Kg/h ($>$9ton per year, $>$90k liters of ink per year) in an industrial system using 70 process cycles and C=100 g/L.


\begin{thebibliography}{99}

\bibitem{Caironi2013} S. Jung, S. D. Hoath, G. D. Martin and I. M. Hutchings, in \textit{Large Area and Flexible Electronics} edited by M. Caironi and Y.-Y. Noh. (Wiley-VCH Verlag GmbH \& Co., 2015).

\bibitem{Leppaniemi2015} J. Lepp\"{a}niemi , O.-H. Huttunen , H. Majumdar and A. Alastalo, Adv. Mater. \textbf{27}, 7168 (2015).

\bibitem{Lau2013} P. H. Lau, K. Takei, C. Wang, Y. Ju, J. Kim, Z. Yu, T. Takahashi, G. Cho and A. Javey, Nano Lett. \textbf{13}, 3864 (2013).

\bibitem{Krebs2010} F. C. Krebs, J. Fyenbob and M. J{\o}rgensen, J. Mater. Chem. \textbf{20}, 8994 (2010).

\bibitem{Dearden2005} A. L. Dearden, P. J. Smith, D.-Y. Shin, N. Reis, B. Derby and P. O'Brien, Macromol. Rapid Commun. \textbf{26}, 315 (2005).

\bibitem{Magdassi2010a}  S. Magdassi, M. Grouchko and A. Kamyshny, Materials \textbf{3}, 4626 (2010).

\bibitem{Wang1997} J. Wang and P. V. A. Pamidi, Anal. Chem. \textbf{69}, 4490 (1997).

\bibitem{Jeong2008} S. Jeong, K. Woo, D. Kim, S. Lim, J. S. Kim, H. Shin, Y. Xia, and J. Moon, Adv. Funct. Mater., \textbf{18}, 679 (2008).

\bibitem{Grouchko2011} M. Grouchko, A. Kamyshny, C. F. Mihailescu, D. F. Anghel and S. Magdassi, ACS Nano \textbf{5}, 3354 (2011).

\bibitem{Lucera2015} L. Lucera, P. Kubis, F. W. Fecher, C. Bronnbauer, M. Turbiez, K. Forberich, T. Ameri, H.-J. Egelhaaf, and C. J. Brabec, Energy Technol. \textbf{3} 373 (2015).

\bibitem{Huang2015}  X. Huang, T. Leng, X. Zhang, J. C. Chen, K. H. Chang, A. K. Geim, K. S. Novoselov and Z. Hu, Appl. Phys. Lett. \textbf{106}, 203105 (2015).

\bibitem{Hosel2013} Markus H¨osel, Roar R. Søndergaard, Dechan Angmo and Frederik C. Krebs, Adv Eng Mater \textbf{15}, 995 (2013).

\bibitem{SommerLarsen2013} P. Sommer-Larsen, M. J{\o}rgensen, R. R. S{\o}ndergaard, M. Hcsel, and F. C. Krebs, Energy Technol. \textbf{1}, 15 (2013).

\bibitem{Krebs2014} F. C. Krebs, N. Espinosa and M. H\"{o}sel, R. R. S{\o}ndergaard and M. J{\o}rgensen, Adv. Mater. \textbf{26}, 29 (2014).

\bibitem{Caironi2015} M. Caironi, Y.-Y. Noh, in \textit{Large Area and Flexible Electronics} (Wiley-VCH Verlag GmbH \& Co. 2015).

\bibitem{Leach2007} J. W. Birkenshaw, in \textit{The Printing Ink Manual, Fifth Edition}, edited by R. H. Leach, R. J. Pierce, E. P. Hickman, M. J. Mackenzie and H. G. Smith. (Springer, The Netherlands, 2007).

\bibitem{Khan2015} S. Khan, L. Lorenzelli and R. S. Dahiya, IEEE Sens J \textbf{15} 3164 (2015).

\bibitem{Tobjork2011} D. Tobj\"{o}rk and R. \"{O}sterbacka, Adv. Mater. \textbf{23}, 1935 (2011).

\bibitem{Hyun2015a} W. J. Hyun, S. Lim, B. Y. Ahn, J. A. Lewis, C. D. Frisbie and L. F. Francis, ACS Appl. Mater. Interfaces \textbf{7}, 12619 (2015)

\bibitem{Merilampi2009} S. Merilampi, T. Laine-Ma and P. Ruuskanen, Microelectron. Reliab. \textbf{49}, 782 (2009)

\bibitem{silverprice} Silverprice http://silverprice.org (accessed Nov 7 2016)

\bibitem{statista} Statista http://statista.com (accessed Nov 7 2016)

\bibitem{Gwent2015} Conductive Carbon Ink C2130925D1 http://gwent.org (accessed Nov 7 2016)

\bibitem{Henkel2015} Henkel Adhesive Technologies, Low-resistivity, screen-printable, carbon ink, LOCTITE EDAG PF 407C E\&C http://henkel-adhesives.com/ (accessed Nov 7 2016)

\bibitem{Dupont2015} DuPont 7102 and BQ242 Conductive Carbon Inks, http://dupont.com (accessed Nov 7 2016)

\bibitem{infomine} Infomine http://infomine.com/investment/metal-prices/copper/5-year/ (accessed Nov 7 2016)

\bibitem{Lachkar1994} A. Lachkar, A. Selmani, E. Sacher, M. Leclerc, R. Mokhliss, Synthetic Met \textbf{66} 209 (1994).

\bibitem{Kim2011} C.-U. Kim, in \textit{Electromigration in thin films and electronic devices, Materials and reliability}, (Woodhead Publishing Limited, 2011)

\bibitem{Rosch2012} Roland R\"{o}sch et al., Energy Environ. Sci. \textbf{5} 6521 (2012).

\bibitem{Lloyd2009} M. T. Lloyd, D. C. Olson, P. Lu, E. Fang, D. L. Moore, M. S. White, M. O. Reese, D. S. Ginleyb and J. W. P. Hsua, J. Mater. Chem. \textbf{19} 7638 (2009).

\bibitem{Sondergaard2014}  R. R. S{\o}ndergaard, N. Espinosa, M. J{\o}rgensen and F. C. Krebs, Energy Environ. Sci., \textbf{7}, 1006 (2014).

\bibitem{Fahmy2009} B. Fahmy and S. A. Cormier, Toxicol. in Vitro, \textbf{23}, 1365 (2009).

\bibitem{Ahamed2010} M. Ahamed, M. A. Siddiqui, M. J. Akhtar, I. Ahmad, A. B. Pant and H. A. Alhadlaq, Biochem. Biophys. Res. Commun. \textbf{396}, 578 (2010).

\bibitem{Karlsson2008} H. L. Karlsson, P. Cronholm, J. Gustafsson and L. M\"{o}ller, Chem. Res. Toxicol., \textbf{21}, 1726 (2008).

\bibitem{Hernandez2008} Y. Hernandez, V. Nicolosi, M. Lotya, F. M. Blighe, Z. Sun, S. De, I. T. McGovern, B. Holland, M. Byrne, Y. K. Gun'Ko, J. J. Boland, P. Niraj, G. Duesberg, S. Krishnamurthy, R. Goodhue, J. Hutchison, V. Scardaci, A. C. Ferrari and J. N. Coleman, Nat. Nanotechnol. \textbf{3}, 563 (2008).

\bibitem{Valles2008} C. Vall\'{e}s, C. Drummond, H. Saadaoui, C. A. Furtado, M. He, O. Roubeau, L. Ortolani, M. Monthioux and A. Pénicaud, J. Am. Chem. Soc. \textbf{130}, 15802 (2008).

\bibitem{Khan2010} U. Khan, A. O'Neill, M. Lotya, S. De, and J. N. Coleman, small \textbf{6}, 864 (2010)

\bibitem{Hasan2010} T. Hasan, F. Torrisi, Z. Sun, D. Popa, V. Nicolosi, G. Privitera, F. Bonaccorso and A. C. Ferrari, Phys. Status Solidi B, \textbf{247} 2953 (2010).

\bibitem{Hernandez2010} Y. Hernandez, M. Lotya, D. Rickard, S. D. Bergin and J. N. Coleman, Langmuir \textbf{26}, 3208 (2010).

\bibitem{Bourlinos2009} A. B. Bourlinos, V. Georgakilas, R. Zboril, T. A. Steriotis and A. K. Stubos, small \textbf{5}, 1841 (2009).

\bibitem{Lotya2009} M. Lotya, Y. Hernandez, P. J. King, R. J. Smith, V. Nicolosi, L. S. Karlsson, F. M. Blighe, S. De, Z. Wang, I. T. McGovern, G. S. Duesberg, and J. N. Coleman, J. Am. Chem. Soc. \textbf{131}, 3611 (2009).

\bibitem{Bonaccorso2012} F. Bonaccorso, A. Lombardo, T. Hasan, Z. Sun, L. Colombo and C. Ferrari, Mater. Today \textbf{15}, 564 (2012).

\bibitem{Torrisi2012} F. Torrisi, T. Hasan , W. Wu , Z.i Sun , A. Lombardo , T. S. Kulmala, G.-W. Hsieh , S. Jung , F. Bonaccorso , P. J. Paul , D. Chu and A. C. Ferrari, ACS Nano \textbf{6}, 2992 (2012).

\bibitem{Martinez2009} J. L. Capelo-Martinez, in \textit{Ultrasound in Chemistry: Analytical Applications} (WILEY-VCH Verlag GmbH, 2009).

\bibitem{Nascentes2001} C. C. Nascentes, M. Korn, C. S. Sousa and M. A. Z. Arruda, J. Braz. Chem. Soc. \textbf{12}, 57 (2001).

\bibitem{Chivate1995} M.M. Chivate and A.B. Pandit, Ultrason. Sonochem. \textbf{2}, 19 (1995).

\bibitem{McClements2005} McClements, in \textit{Food Emulsions Principles, Practices, and Techniques} (CRC Press, 2005).

\bibitem{Secor2013} E. B. Secor, P. L. Prabhumirashi, K. Punbekar, M.l L. Geier and M. C. Hersam, J. Phys. Chem. Lett. \textbf{4}, 1347 (2013).

\bibitem{Hyun2015b} W. J. Hyun, E. B. Secor, M. C. Hersam , C. D. Frisbie and L. F. Francis, Adv. Mater. \textbf{27}, 109 (2015).

\bibitem{Paton2014}  K. R. Paton et al. Nat. Mater. \textbf{13}, 624 (2014).

\bibitem{Brookfield}  Brookfield Engineering - Viscosity Glossary, http://www.brookfieldengineering.com/education
/viscosity$\_$glossary.asp (accessed Nov 7 2016)

\bibitem{Paul2004} E. L. Paul, V. A. Atiemo-Obeng and S. M. Kresta, in \textit{Hanbook of industrial mixing, Science and Practice} (John Wiley \& Sons, 2004).

\bibitem{Wang2011} J. Wang, K. K. Manga, Q. Bao and K. P. Loh, J. Am. Chem. Soc. \textbf{133}, 8888 (2011).

\bibitem{Panagiotou2008b}  T. Panagiotou, S.V. Mesite, J.M. Bernard, K.J. Chomistek and R.J. Fisher, NSTI-Nanotech, \textbf{1}, 688 (2008).

\bibitem{Posch2008} A. Posch, in \textit{2D PAGE. Volume 1: Sample preparation \& pre-fractionation, Methods in molecular biology} (Humana Press, 2008).

\bibitem{microfluidicscorp} http://www.microfluidicscorp.com/

\bibitem{Lajunen2014} T. Lajunen, K. Hisazumi, T. Kanazawa, H. Okada, Y. Seta, M. Yliperttula, A. Urtti and Y. Takashima, Eur J Pharm Sci \textbf{62}, 23 (2014).

\bibitem{Tang2013} S. Y. Tang, P. Shridharan and M. Sivakumar, Ultrason Sonochem \textbf{20}, 485 (2013).

\bibitem{Jafari2007} S. M. Jafari, Y. He, B. Bhandari, J Food Eng \textbf{82}, 478 (2007).

\bibitem{Panagiotou2008} T. Panagiotou, J. M. Bernard and S. V. Mesite, NSTI-Nanotech \textbf{1}, 39 (2008).

\bibitem{Benatto2014} G. A. dos Reis Benatto, B. Roth, M. V. Madsen , M. H\"{o}sel , R. R. S{\o}ndergaard, M. J{\o}rgensen and F. C. Krebs, Adv. Energy Mater. \textbf{4}, 1400732 (2014).

\bibitem{Nisato2016} G. Nisato, D. Lupo and S. Ganz, in \textit{Organic and Printed Electronics: Fundamentals and Applications} (CRC Press, 2016).

\bibitem{Imerys} Imerys http://www.imerys-graphite-and-carbon.com/ (accessed Nov 7 2016)

\bibitem{Kouroupis-Agalou2014} K. Kouroupis-Agalou, A. Liscio, E. Treossi, L. Ortolani, V. Morandi, N. M. Pugno and V. Palermo, Nanoscale \textbf{6} 5926 (2014).

\bibitem{Moeck2011} P. Moeck, S. Rouvimov, E. F. Rauch, M. V\'{e}ron, H. Kirmse, I. H\"{a}usler, W. Neumann, D. Bultreys, Y. Maniette and S. Nicolopoulos, Cryst. Res. Technol. \textbf{46} 589 (2011).

\bibitem{nanomegas} Nanomegas http://nanomegas.com (accessed Nov 7 2016)

\bibitem{Gammer2015} C. Gammer, V. Burak-Ozdol, C. H. Liebscher and A. Minor, Ultramicroscopy, \textbf{155}, 1 (2015).

\bibitem{Punckt2013} C. Punckt, F. Muckel, S. Wolff, I. A. Aksay, C. A. Chavarin, G. Bacher, and W. Mertin, Appl. Phys. Lett. \textbf{102}, 023114 (2013).

\bibitem{ASTM1} ASTM E1131-08 Standard Test Method for Compositional Analysis by Thermogravimetry (West Conshohocken, PA, 2014).

\bibitem{Shirley1972} D. A. Shirley, Phys. Rev. B \textbf{5}, 4709 (1972).

\bibitem{Welham1998} N.J. Welham, J.S. Williams, Carbon \textbf{36}, 1309 (1998).

\bibitem{Benson2014} J. Benson, Q. Xu, P. Wang, Y. Shen, L. Sun, T. Wang, M. Li and P. Papakonstantinou, ACS Appl. Mater. Interfaces \textbf{6}, 19726 (2014).

\bibitem{ASTM2} ASTM E1078-14 Standard Guide for Specimen Preparation and Mounting in Surface Analysis (West Conshohocken, PA, 2014).

\bibitem{Moulder1992} J. F. Moulder, W. F. Stickle, P. E. Sobol, and K. Bomben,  \textit{Handbook of X-Ray Photoelectron Spectroscopy} (Physical Electronics Division, Perkin-Elmer Corporation, 1992).

\bibitem{Phaner-Goutorbe1994} M. Phaner-Goutorbe, A. Sartre, and L. Porte, Microsc. Microanal. Microstruct. \textbf{5}, 283 (1994).

\bibitem{Briggs1990} D. Briggs and M. P. Seah,  \textit{Practical Surface Analysis, Auger and X-Ray Photoelectron Spectroscopy} (Wiley, 1990).

\bibitem{Yang2009} D. Yang, A. Velamakannia, G. Bozoklu, S. Park, M. Stoller, R. D. Piner, S. Stankovich, I. Jung, D. A. Field, C. A. Ventrice Jr., R. S. Ruoff, Carbon \textbf{47}, 145 (2009).

\bibitem{Drewniak2015} S. Drewniak, R. Muzyka, A. Stolarczyk, T. Pustelny, M. Kotyczka-Morańska, and M. Setkiewicz, Sensors (Basel) \textbf{16}, 103 (2015).

\bibitem{Haubner2010} K. Haubner, J. Murawski, P. Olk, L. M. Eng, C. Ziegler, B. Adolphi, and E. Jaehne, Chemphyschem \textbf{11}, 2131 (2010).

\bibitem{Tuinstra1970} F. Tuinstra and J. L. Koenig, J. Chem. Phys. \textbf{53}, 1126 (1970).

\bibitem{Ferrari2000} A. C. Ferrari and J. Robertson, Phys. Rev. B \textbf{61}, 14095 (2000).

\bibitem{Thomsen2000} C. Thomsen and S. Reich, Phys. Rev. Lett. \textbf{85}, 5214 (2000).

\bibitem{Baranov1987} A. V. Baranov, A. N. Bekhterev, Ya. S. Bobovich and  V. I. Petrov, Opt. Spectroscopy  \textbf{62}, 612 (1987).

\bibitem{Pocsik1998} I. Pocsik, M. Hundhausen, M. Koos and L. Ley, J. Non-Cryst. Solids  \textbf{227}, 1083 (1998).

\bibitem{Piscanec2004} S. Piscanec, M. Lazzeri, F. Mauri, A. C. Ferrari and J. Robertson, Phys. Rev. Lett. \textbf{93}, 185503 (2004).

\bibitem{Ferrari2006} A. C. Ferrari, J. C. Meyer, V. Scardaci, C. Casiraghi, M. Lazzeri, F. Mauri, S. Piscanec, D. Jiang, K. S. Novoselov, S. Roth and A. K. Geim, Phys. Rev. Lett. \textbf{97}, 187401 (2006).

\bibitem{Basko2009} D. M. Basko, S. Piscanec, A. C. Ferrari, Phys. Rev. B \textbf{80}, 165413 (2009).

\bibitem{Ferrari2013} A. C. Ferrari and D. M. Basko, Nat Nano \textbf{8}, 235 (2013).

\bibitem{Ferrari2001} A. C. Ferrari and J. Robertson, Phys. Rev. B \textbf{64}, 075414 (2001).

\bibitem{Ferrari2003} A. C. Ferrari, S. E. Rodil and J. Robertson, Phys. Rev. B \textbf{67}, 155306 (2003).

\bibitem{Ferrari2007} A. C. Ferrari, Solid State Commun \textbf{143}, 47 (2007).

\bibitem{Ummartyotin2015} S. Ummartyotin and H. Manuspiya, Renew Sust Energ Rev \textbf{41}, 402 (2015).

\bibitem{Risio2007} S. Di Risio, and N. Yan, Macromol. Rapid Commun. \textbf{28}, 1934 (2007).

\bibitem{Pavinatto2015} F. J. Pavinatto, C. W. A. Paschoal and A. C. Arias, ‎Biosens. Bioelectron \textbf{67}, 553 (2015).

\bibitem{Mezger2006} T. G. Mezger,  \textit{The Rheology Handbook: For Users of Rotational and Oscillatory Rheometers} (Vincentz Network GmbH \& Co KG, 2006).

\bibitem{Lin2008} H.-W. Lin, C.-P. Chang, W.-H. Hwu, M.-D. Ger, J. Mater. Process. Technol. \textbf{197}, 284 (2008).

\bibitem{Benchabane2008} A. Benchabane, K. Bekkour, Colloid Polym Sci \textbf{286}, 1173 (2008).

\bibitem{Nijenhuis2007} K. te Nijenhuis, G. H. McKinley, S. Spiegelberg, H. A. Barnes, N. Aksel, L. Heymann, J. A. Odell,  \textit{Springer Handbook of Experimental Fluid Mechanics}, edited by C. Tropea, A. L. Yarin, J. F. Foss (Springer-Verlag Berlin Heidelberg, 2007).

\bibitem{Irgens2014} F. Irgens, \textit{Rheology and Non-Newtonian Fluids} (Springer Verlag, Cham, Heidelberg et al. 2014).

\bibitem{deButts1957} E. H. deButts, J. A. Hudy and J. H. Elliott, Ind. Eng. Chem. \textbf{49}, 94 (1957).

\bibitem{Elliot1974} J. H. Elliot and A. J. Ganz, Rheologica Acta \textbf{13}, 670 (1974).

\bibitem{Clasen2001} C. Clasen, W.-M. Kulicke, Progress in Polymer Science \textbf{26}, 1839 (2001).

\bibitem{Steffe1996} J. F. Steffe,  \textit{Rheological Methods in Food Process Engineering}, (Freeman Press, 2807 Still Valley Dr. East Lansing, MI 48823, USA, 0-9632036-1-4, 1996).

\bibitem{Roff1971} W. J. Roff,  J. R. Scott., \textit{Fibres, Films, Plastics and Rubbers: A Handbook of Common Polymers}, (Butterworths, 1971).

\bibitem{Xia1988} W. Xia, M. F. Thorpe, Phys. Rev. A. \textbf{38}, 2650 (1988).

\bibitem{Li2009} W. Li, B. Sun, P. Wu, Carbohydr. Polym. \textbf{78}, 454 (2009).

\bibitem{audioposter} Audioposter, http://www.audioposter.com/ (accessed Nov 7 2016)

\bibitem{Munson2009} B. R. Munson, D. F. Young, T. H. Okiishi, W. W. Huebsch,  \textit{Fundamentals of Fluid Mechanics, Sixth Edition} (John Wiley \& Sons, Inc., 2009).

\bibitem{Rouse1946} H. Rouse, Elementary mechanics of fluids, (Dover Publications Inc. New York, 1946)

\bibitem{Secco2014} R. A. Secco, M. Kostic, J. R. deBruyn, Measurement, Instrumentation, and Sensors Handbook, Second Edition: Spatial, Mechanical, Thermal, and Radiation Measurement, Edited by J. G. Webster, H. Eren, (CRC Press Taylor \& Francis Group, 2014).

\bibitem{Holman1986} J. P. Holman, in \textit{Heat Transfer} (McGraw-Hill. 1986).

\bibitem{Moody1944} L. F. Moody and N. J. Princeton, Transactions of the ASME \textbf{66} 671 (1944).

\bibitem{Siddiqui2009}S. W. Siddiqui, Y. Zhao, A. Kukukova and S. M. Kresta, Ind. Eng. Chem. Res. \textbf{48}, 7945 (2009).

\bibitem{Boxall2012} J. A. Boxall, C. A. Koh, E. D. Sloan, A. K. Sum, D. T. Wu, Langmuir \textbf{28}, 104 (2012).

\bibitem{Zhang2012} J. Zhang, S. Xu and W. Li, Chem. Eng. Process \textbf{57} 25 (2012).

\bibitem{Kolmogorov1941} A. N. Kolmogorov, Dokl. Akad. Nauk SSSR \textbf{30} 299 (1941).

\bibitem{Richardson1922} L. F. Richardson, Weather prediction by numerical process, Cambridge University Press \textbf{48} 282 (1922).

\bibitem{Varrla2014}  E. Varrla, K. R. Paton, C. Backes, A. Harvey,  R. J. Smith, J. McCauley, J. and N. Coleman, Nanoscale \textbf{6}, 11810 (2014).

\end{thebibliography}
\end{document}